\documentclass[sigconf]{acmart}

\AtBeginDocument{%
  \providecommand\BibTeX{{%
    \normalfont B\kern-0.5em{\scshape i\kern-0.25em b}\kern-0.8em\TeX}}}

\setcopyright{acmcopyright}
\copyrightyear{2018}
\acmYear{2018}
\acmDOI{XXXXXXX.XXXXXXX}

\acmConference{}{}{}

\acmISBN{978-1-4503-XXXX-X/18/06}

\usepackage[utf8]{inputenc}
\usepackage{xcolor}
\usepackage{hyperref}
\usepackage{xspace}
\usepackage{color}
\usepackage{listings}
\usepackage{acronym}
\usepackage[T1]{fontenc}
\usepackage{lmodern}
\usepackage{multirow}
\usepackage{colortbl}
\usepackage{todonotes}
\usepackage{url}
\usepackage[export]{adjustbox}
\usepackage{tcolorbox}
\usepackage{comment}
\usepackage{colortbl}
\usepackage{tcolorbox}
\usepackage{lipsum}
\usepackage{multicol}
\usepackage{amsmath}
\usepackage{float}
\usepackage{graphicx}
\usepackage{algpseudocode}
\usepackage{algorithm}
\usepackage{listings}
\usepackage[flushleft]{threeparttable}
\usepackage{subcaption}%

\definecolor{lightgray}{rgb}{.9,.9,.9}
\definecolor{mauve}{rgb}{0.58,0,0.82}
\definecolor{darkgray}{rgb}{.4,.4,.4}
\definecolor{darkgreen}{rgb}{0, 0.39, 0.00}
\definecolor{brinkpink}{rgb}{0.98, 0.38, 0.5}
\definecolor{brightpink}{rgb}{1.0, 0.0, 0.5}
\definecolor{Gray}{gray}{0.7}
\acrodef{SCD}{Simple Circuit Description}
\acrodef{ABE}{Attribute-Based Encryption}
\acrodef{LWE}{Learning With Errors}
\acrodef{GC}{Garbled Circuit}
\acrodef{SFE}{Secure Function Evaluation}
\acrodef{ECU}{Electronic Control Units}
\acrodef{CAN}{Controller Area Network}
\acrodef{DoS}{Denial of Service}
\acrodef{PIds}{Parameter Identifiers}
\acrodef{V2V}{Vehicle to Vehicle}
\acrodef{OBD}{On-Board Diagnostics}
\acrodef{DNN}{Deep Neural Network}
\acrodef {RNN}{Recurrent Neural Network}
\acrodef{LSTM RNN}{Long Short-Term Memory Recurrent Neural Network}
\acrodef{CI/CD}{continuous integration, delivery, and deployment systems}
\acrodef{K8s}{Kubernetes}
\acrodef{SSCS}{software supply chain system}
\acrodef{CI}{Continuous Integration}
\acrodef{VM}{Virtual Machine} 
\acrodef{CRUD}{create, read, update, and delete}

\usepackage{amsfonts}

\begin{document}

\title{Making Secure Software Insecure without Changing Its Code: The Possibilities and Impacts of Attacks on the DevOps Pipeline}

\author{Nicholas Pecka}
\affiliation{%
  \institution{Iowa State University}\country{USA}}

\author{Lotfi ben Othmane}
\affiliation{%
  \institution{Iowa State University}\country{USA}}

\author{Altaz Valani}
\affiliation{%
 \institution{Security Compass}\country{Canada}}

\renewcommand{\shortauthors}{Pecka  et al.}

\begin{abstract}

Companies are misled into thinking they solve their security issues by using a DevSecOps system. This paper aims to answer the question: Could a DevOps pipeline be misused to transform a securely developed application into an insecure one? To answer the question, we designed a typical DevOps pipeline utilizing \ac{K8s} as a case study environment and analyzed the applicable threats. Then, we developed four attack scenarios against the case study environment: maliciously abusing the user's privilege of deploying containers within the K8s cluster, abusing the Jenkins instance to modify files during the \ac{CI/CD} build phase, modifying the \ac{K8s} DNS layer to expose an internal IP to external traffic, and elevating privileges from an account with \ac{CRUD} privileges to root privileges. The attacks answer the research question positively: companies should design and use a secure DevOps pipeline and not expect that using a DevSecOps environment alone is sufficient to deliver secure software.

\end{abstract}



\maketitle

\section{Introduction}\label{sec:intro}

Companies are adopting the DevOps paradigm~\cite{DevOps2021} where development and operation teams coexist and focus on a consistent development and delivery process. DevSecOps~\cite{8543383,7784617,DevSecOps2022}, a section of DevOps, incorporates the benefits that DevOps has brought, and includes a security mindset. This mindset helps to explicitly and intentionally evolve security maturity, without assuming that the system under development is secure on the sole basis that DevOps principles are being followed.

\begin{figure}
\centering
		\includegraphics[width=0.48\textwidth]{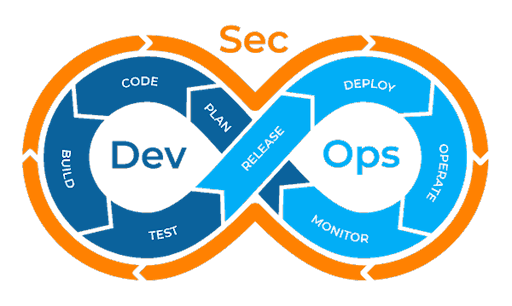}
		\caption{DevSecOps Model~\cite{CICDMODEL2021}.}
		\label{fig:cicdmodel}
		\vspace{-0.25in}
\end{figure}

To further evolve the DevSecOps mentality, companies have been adopting microservice architectures~\cite{Microservices2021} as their vision closely aligns with DevSecOps ideals around focused, incremental development. DevSecOps, as depicted by Figure~\ref{fig:cicdmodel}, is typically used to deploy complex software as a set of applications. This paper focuses on the deployment of a DevOps pipeline, which includes a code repository (e.g., GitHub, GitLab), a \ac{CI/CD} system~\cite{CICD2021} (Jenkins), a repository for storing packaged images (DockerHub, Quay.io, etc.) and the underlying architecture serving the components~\ac{K8s}~\cite{Kub2021,automatedCloudInfrastructure2021,ContinuousIntegrationAWS2021}. These were determined based on their close alignment with a DevSecOps mentality. For example, K8s allows a system to be better scaled, monitored, and maintained, including systems that use machine learning~\cite{DevOpsContinousAutomation2021}.
        
The integration of development and production environments encouraged companies to adopt DevSecOps by integrating secure software practices and activities into their DevOps systems. Companies adopt DevSecOps models believing it solves their security issues without additional input~\cite{7784617}. The automation of DevSecOps systems including the use of K8s, might, however, introduce new security threat vectors despite the great deal of benefits it provides. The question is: \emph{Could a DevOps pipeline be misused to transform a securely developed application into an insecure one?} Companies are being misled into thinking they have solved their security issues simply by utilizing a DevSecOps process. Answering the question positively would raise awareness about the need to use secure software supply chain systems.



To answer the research question, we created (as a case study) a DevOps environment that uses a K8s environment and implements various components needed for the application development life cycle. Then, we performed threat modeling~\cite{ThreatModeling2021} of the system. Threat modeling is a process where a system is analyzed for potential security attacks that take advantage of vulnerabilities, quantifying threats, and recommending appropriate remediation~\cite{ThreatModeling2021}. Along with those objectives, we aimed to acquire knowledge on pre-existing vulnerabilities and also potential areas we could exploit for testing. From that research we derived four attack scenarios: (1) retrieval of application data utilizing a custom app that leverages the K8s DNS, (2) manipulate the CI/CD application Jenkins and install a backdoor, (3) expose an internal cluster IP to external, and (4) leverage a hostPath volume to escape a namespace and gain root access on the host. We specifically look at the concept of privilege escalation throughout these scenarios.

The contributions of the paper are:
\begin{enumerate}
\item Developing a threat model of a DevOps environment utilizing Strimzi application as a case study.
\item Designing four attacks scenarios that demonstrate four of the threats to the DevOps environment.
\item Proposing mitigation techniques for the identified threats.
\end{enumerate}

The tests show that DevOps model weaknesses could create insecure \ac{SSCS}. The resources for the project including the attack videos are shared at Ref.~\cite{Pecka2022}.

The remainder of this paper is organized as follows: Section~\ref{sec:relworks} discusses related work concerning the structuring of DevSecOps systems and their security, Section~\ref{sec:casestudy} provides information on the experimentation environment along with the components that will make up that environment, Section~\ref{sec:attacks} describes the penetration tests performed to test our hypothesis, Section~\ref{sec:protections} discusses mitigation against the penetration tests performed in Section~\ref{sec:attacks}, and Section~\ref{sec:conclusions} describes the future work and concludes the paper.

\section{Related work}\label{sec:relworks}

Understanding potential entry points to thwart attackers is vital information. Shamim et al.~outlined and explained in Ref~\cite{CommandmentsOfKubernetes2020} the multiple levels of security including authentication, security policies, logging, network isolation, encryption, patching, SSL/TLS, and others in great details. They derived their findings from over 100 internet artifacts. The individual items outlined are not a comprehensive list but were found to be the most affected points of entry across the examined artifacts. Minna et al. extended Shamim etal.'s work ~\cite{UnderstandingSecurityKubernetes2021} by outlining various network-security issues pertaining to a \ac{K8s} cluster, such as Pod netns by a Pause Container, CNI Plug-Ins Jeopardy, software isolation of resources, network policies limitations, multi tenant K8s clusters, dynamic nature of K8s objects, virtual network infrastructure, and not embedded distributed tracing. In addition, the authors mapped the security of K8s to the Microsoft K8s threat matrix~\cite{kubeMatrix2021}. Karamitsos et al~\cite{DevOpsContinousAutomation2021} discuss the impact of business manager in deciding on accepting risks of the DevOps systems due to associated time and cost. 

Bertucio analyzed the security of \acf{SSCS}~\cite{GoogleHypothesis2021}. They break down each component of the supply chain and provide a risk and remediation of each section. They outline a SSCS to depict the points highlighted throughout the blog provided by google. The paper looks further in detail about specific elements of the supply chain and provides real world examples from an attackers point of view followed by mitigation's to said attacks.

For a further look into various attack scenarios, a github user by the name of madhuakula has created an interactive playground called Kubernetes Goat~\cite{K8SGOAT2021}. Users can either follow the instructions on the page to setup their own vulnerable \ac{K8s} environment or use the built in interactive playground to follow various penetration testing scenarios.

On a related topic, attackers realized that public \ac{CI} platforms are resource-rich but loosely protected free Internet services and started exploiting that for Cryptomining. For instance, Li et al.\cite{LLCW2022} discovered 1,974 Cijacking instances, 30 campaigns across 12 different cryptocurrencies on 11 mainstream CI platforms. Further, they unveils the evolution of cryptojacking attack strategies on the \ac{CI} platforms in response to the protection put in place by these platforms, the duration of the mining jobs (as long as 33 months), and their life cycle. They also discovered that the revenue of the attack is over \$20,000 per month.

\begin{figure}
\centering
		\includegraphics[width=0.48\textwidth]{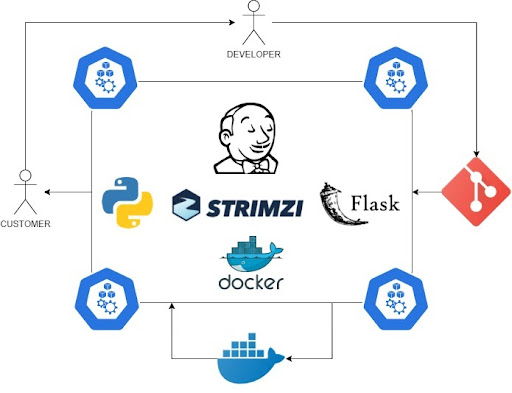}
		\caption{The case study: the implemented DevOps environment.}
		\label{fig:devopspipeline}
			\vspace{-0.2in}
\end{figure}

\begin{table}[bt]
        \centering
          \caption{Partial threat model of DevOps pipeline case study.}
\label{tab:threatModelComp}  
\begin{tabular}{|p{0.5in}|p{0.6in}|p{0.9in}|p{0.8in}| }

\hline 
\rowcolor{gray!18}
 \begin{center}
\textbf{{\small System}}
\end{center}
& \begin{center}
\textbf{{\small Potential Flaw}}
\end{center}
 & \begin{center}
\textbf{{\small Threat Description}}
\end{center}
 & \begin{center}
\textbf{{\small Mitigation}}
\end{center}
 \\
\hline 
Git& Account Compromise& Potential privileged account compromise& Enable 2 factor auth\\
\hline 
Jenkins& Control over CI/CD& Account can be used to alter build configs& Enable 2 factor auth\\
\hline
Docker& Docker Pull& Potential infected container& Implement vuln scanning for containers\\
\hline
\ac{K8s}& DDOS& Front end apps externally exposed& Proper security checks (ex. check in place to prevent multiple auth)\\
\hline
Strimzi& Network Policy& Default pods can access listeners& Configure proper network policy\\
\hline
Custom App& Non sanitized fields& Potential for malicious input& Patch to allow sanitized\\
\hline
\multicolumn{3}{l}{} \\

\end{tabular}%
	\vspace{-0.25in}
\end{table}

\begin{table*}[bt] 
        \centering
          \caption{Outline of the attack scenarios.}
\label{tab:attackscenarios}  
\begin{tabular}{|p{0.15in}|p{1.2 in}|p{2.9 in}|p{.7 in}|p{1.3 in}| }

\hline 
\rowcolor{gray!18}
 \begin{center}
\textbf{{\small ID}}
\end{center}
& \begin{center}
\textbf{{\small Prerequisite}}
\end{center}
& \begin{center}
\textbf{{\small Attack}}
\end{center}
 & \begin{center}
\textbf{{\small Component}}
\end{center}
& \begin{center}
\textbf{{\small Attacker Gains}}
\end{center}
 \\
\hline
1& Ability to deploy applications within the same namespace as target&Deploys malicious application, leverages application to siphon data from target application&Strimzi&Potential to compromise sensitive data within kafka\\
\hline
2& Access to privileged Jenkins account&Authenticates to Jenkins, Edits build job to input bad payload, once deployed accesses application through backdoor created from bad payload&Jenkins&Installation of potential backdoor into application being built\\
\hline
3& Account with privileges to add ingress objects to desired target&Authenticates to K8s cluster, edits the networking of target applications service, inputs an ingress object to expose externally&K8s Networking&Ability to connect to app from external\\
\hline
4& Account with CRUD privileges&Deploys application in namespace with hostPath volume, leverages the volume by rooting to the host system, gains K8s credentials and has full access to cluster&K8s storage, Docker, Worker Node&Root access to Host box along with K8s cluster\\
\hline
\multicolumn{4}{l}{} \\

\end{tabular}%
	\vspace{-0.25in}
\end{table*}

\section{Case study setup}\label{sec:experiment}\label{sec:casestudy}

This section describes a system that we setup to demonstrate the use of a \ac{SSCS} to integrate and deploy a secure application (Strimzi) as an insecure software. 

We developed a simple Web application to demonstrate the use of an application with certain privileges to access other components of the DevOps environment~\cite{pyProducerConsumer2021}. The application is developed in Python~\cite{Python2021} utilizing the micro web framework Flask~\cite{Flask2021} to provide the user/attacker a front-end UI when deployed to the \ac{K8s} cluster. The application uses the Apache Kafka\cite{Kafka2021} library to interact with Strimzi~\cite{Strimzi2021}. Strimzi is used to streamline the deployment of Apache Kafka~\cite{Kafka2021} on \ac{K8s}. This application will serve as our secure application for our later attack scenarios. 

ESXi~\cite{ESXI2021} is used as the hypervisor~\cite{Hypervisor2021} to manage four \acp{VM}'s that host the experimentation environment. ESXi was built on a bare metal server with an i7-6700K CPU @ 4.00GHz, 4 CPU cores, and 32GB RAM. GitHub~\cite{GitHub2021} is used as a code repository for the application. GitHub serves as the trigger point for the Jenkins~\cite{Jenkins2021} job. Jenkins is an industry standard for CI/CD~\cite{CICD2021}. 

The application components are deployed to K8s ~\cite{Kub2021}, an orchestration tool for docker~\cite{Docker2021} containers that allows a collection of containers to be monitored, managed, and sized at scale. The kube-apiserver is leveraged for the scheduler, controller-manager, and the etcd components to communicate so they can exist separately allowing them to be decoupled. There is then a kubelet on each of the worker nodes that will call back to the apiserver so the other components can manage the cluster properly. An important note is that K8s shrouds the containers in its own networking layer. \ac{K8s} aligns closely with the DevSecOps mentality and these functionalities provide DevSecOps great tooling. Due to this, however, it might be assumed the applications within a \ac{K8s} cluster are secure. This paper aims to prove that insider attackers can exploit security weaknesses of DevSecOps pipeline to transform a secure software into an insecure one. 

Figure~\ref{fig:devopspipeline} depicts the DevOps pipline case study. From developer to the customer, a developer begins by submitting their code commit to GitHub. A webhook listener from Jenkins triggers from the latest code push to Github and initiates the corresponding Jenkins build job. During the build job, Jenkins pulls the source repository from GitHub including the new code changes. Next, it compiles and packages everything into a docker container utilizing built in docker functions to prepare the application for future deployment. Once built, the docker container is pushed to DockerHub~\cite{DockerHub2021} and tagged with the build version. Jenkins proceeds to log into the K8s cluster and utilizes docker commands to download the docker image on DockerHub onto the K8s cluster. The image is then deployed to the K8s cluster. The customer then evaluates the new version and provides user feedback to the development team to complete the DevOps system.

\begin{lstlisting}[numbers=left, frame=single,caption={Steps to perform the Retrieve Information in Topic attack.},captionpos=b, label={lst:infoTopic}]
// Authenticate to Kubernetes cluster
Confirm kubeconfig matches the desired K8s cluster

// Deploy Strimzi
kubectl apply -f name_of_strimzi.yaml

// Deploy Custom Application
kubectl apply -f name_of_custom_app.yaml

// Verify the applications are in ready state
kubectl get pods \
-n name_of_namespace_where_apps_are_located

// Locate Strimzi internal clusterIP
kubectl get services \
-n name_of_namespace_where_strimzi_is_located

// Locate URL of custom application
kubectl get services \
-n name_of_namespace_where_custom_app_is_located

// Navigate to URL from previous step

// Populate fields of custom application to
// connect to Strimzi
Plug in values clusterIP:Port of Strimzi

// Verify data is sent/received to/form Strimzi
Data will be displayed after executing command

\end{lstlisting}

\section{Attack demonstration scenarios on the DevOps case study}\label{sec:attacks}

The components within the research environment described in Section~\ref{sec:casestudy} were broken down and analyzed for their inputs and outputs. Table~\ref{tab:threatModelComp} outlines each component of the DevOps pipeline by system, one potential flaw example, associated threat, and example of mitigation. Through this, we derive potential threats to be used against the system. Table~\ref{tab:attackscenarios} outlines the attacks in the form of prerequisite, attack description, affected component, and attacker's gains from the attack.

This section describes four attacks against the case study environment: maliciously abusing the user's privilege of deploying containers within the K8s cluster, abusing the Jenkins instance to modify files during the CI/CD build phase, modifying the \ac{K8s} DNS layer to expose an internal IP to external traffic, and elevate privileges from a \ac{CRUD} privileged account to a root privileged account.

\subsection{Retrieve Information in Topic}

The first attack deals with an attacker having privileges to deploy containers to a \ac{K8s} cluster. The goal is to compromise information from secure applications within the \ac{SSCS} that exist within the cluster. This data is generally only available within the \ac{K8s} cluster itself due to the K8s DNS layer. The attacker creates a custom application that, when deployed, will provide a front facing web UI. The custom app used was created as part of this research and is available at Ref.~\cite{pyProducerConsumer2021}. Once deployed within the cluster, the attacker connects to the front facing UI of the new application and leverage the application to compromise data by siphoning data from the Strimzi application using the custom application. This could lead to secretive data being compromised depending on what the custom application is able to retrieve. Listing~\ref{lst:infoTopic} lists the commands to produce the attack.

\begin{lstlisting}[numbers=left, frame=single,caption={Steps to perform the Manipulate CI/CD to modify files attack.},captionpos=b, label={lst:cicdModification}]
// Authenticate to Jenkins
Jenkins login - www.name_of_jenkins_url.com
input username/password

// Locate Build Step
Navigate to proper build step

// Modify Build Step
Select build step _for modification

// Edit build step by inputting malicious payload
malicious_payload_code

// Execute build job
Run Jenkins build

// Authenticate to K8s cluster
Confirm kubeconfig matches desired K8s cluster

// Pull newly created malicious docker image
docker pull repo_name/image_name/tag

// Deploy malicious docker image
kubectl apply -f name_of_image.yaml

// Trigger malicious payload
Perform triggering action
\end{lstlisting}

\vspace{-0.3in}

\subsection{Manipulate CI/CD by Modifying the Files}

The second attack deals with an attacker that has privileges to the Jenkins instance serving the \ac{CI/CD}. The attacker selects a build step and then modifies the files prior to being packaged and deployed to an online container repository, as in listing~\ref{lst:cicdModification}. Once the modified application is deployed, anything may trigger the malicious payload. This could put multiple systems at risk if the application modified is heavily used across a wide array of organizations (open source application such as Strimzi is a great example).

\subsection{Kubernetes Expose clusterIP to External}

The third attack deals with an attacker that has access to the K8s cluster to manipulate networking protocols. K8s provides internal cluster IPs, nodeports, and other ingress type objects for \ac{K8s} resources. These objects allow for internal applications to communicate across the cluster, and to external sources. Services in K8s start with an internal cluster IP that allows for communication with other services within the \ac{K8s} cluster. The attacker can expose the cluster IP with another ingress object such as a nodeport. The nodeport will attach an external URL that will allow external applications to contact the internal K8s application via the nodeport. With this, an attacker can hook directly into the now insecure application (courtesy of the recent \ac{K8s} configuration) and siphon secretive data. Listing~\ref{lst:clusterIPExpose} lists the commands to produce the attack.

\begin{lstlisting}[numbers=left, frame=single,caption={Steps to perform the Kubernetes expose clusterIP to external attack.},captionpos=b, label={lst:clusterIPExpose}]
// Authenticate to K8s cluster
Confirm kubeconfig matches desired K8s cluster

// Deploy Strimzi
kubectl apply -f name_of_strimzi.yaml

// Verify Strimzi is in ready state
kubectl get pods -n namespace_of_Strimzi

// Locate Strimzi internal clusterIP
kubectl get services -n namespace_of_Strimzi

// Add NodePort network object to Strimzi
kind: Service
apiVersion: v1
metadata:
  name: strimzi-service
spec:
  selector:
    app: strimzi_app
  ports:
  - protocol: TCP
    port: Strimzi_Port
    nodePort: (30000-32767) - # in _this range
  type: NodePort
  
// Verify Strimzi is exposed
Contact newly exposed IP

\end{lstlisting}

\vspace{-0.25in}

\subsection{Kubernetes hostPath Namespace Breakout}
 
The final attack deals with a hostPath namespace breakout~\cite{namespaceBreakout2021}. A namespace in K8s allows for network segregation and to map deployments too when created. An attacker requires access to a service account with CRUD (create, retrieve, update, delete) privileges in any namespace within the cluster. The attacker deploys a pod within the allowed namespace, then proceeds to abuse the hostPath volume to mount an escape for privilege escalation. A hostPath volume is a storage object that mounts a file or directory from the host node's file system into the pod. Once the attacker deploys their malicious pod, they then exec into it and chroot to access the node's root file system due to the hostPath volume mount exploit. The attacker can then find the kubeconfig files on the host and gain cluster admin privileges. Through this, they can target our secure applications that may exist on other nodes and perform malicious actions against them including editing, deletion, and more. Listing~\ref{lst:nsBreakout} lists the commands to produce the attack.

\begin{lstlisting}[numbers=left, language=C, frame=single,caption={Steps to perform the Kubernetes Namespace Breakout attack.},captionpos=b, label={lst:nsBreakout}]
// Authenticate to K8s cluster
Confirm kubeconfig matches desired K8s cluster

// Ensure service account has CRUD privileges
// Launch attacker pod with hostPath volume attached
kubectl apply -f attacker_pod_name.yaml

// Exec into the attacker pod
kubectl -n crud_namespace \
exec -it attack_pod_name bash

// Verify account level is not admin
kubectl get secrets -n kube-system

// Verify pod creation in developer ns
Kubectl auth can-i create pod -n crud_namespace

// Check where at in host
echo ${uname -n}

// Execute command to escalate privileges
chroot /host/ bash

// Verify location by checking running containers
docker ps

// Locate kubecfg files and view K8s cluster
/location/to/kubectl \
--kubeconfig=/location/to/kubecfg-kube-node.yaml

// Check pods within K8s cluster
/location/to/kubectl \
--kubeconfig=/location/to/kubecfg-kube-node.yaml \
get pods -A

// Delete a pod within the K8s cluster
/location/to/kubectl \
--kubeconfig=/location/to/kubecfg-kube-node.yaml \
delete pod pod_name -n pod_namespace


\end{lstlisting}
 \vspace{-0.25in}
 
\section{Proposed protection mechanisms}\label{sec:protections}

The common theme of the attacks is privilege escalation. Thus, the first protection from the attacks is use of the principle of least privilege when managing a SSCS within the utilized DevSecOps model \ac{K8s}; limiting account access shrinks the attack vector, as attackers will have less victims to choose from, to perform the type of attacks described in this paper. This section proposes protection mechanisms against the reported attack scenarios of section~\ref{sec:attacks}. 

\noindent{\bf Protection from deploying malicious application.} The main protection from deploying malicious applications and disclosing confidential information is to limit users privileges. We recommend implementing service accounts that are tied to specific namespaces to prevent users from deploying containers outside their dedicated area.

\noindent{\bf Protection from the CI/CD manipulation.} The main protection from manipulating the CI/CD pipeline is to restrict access to Jenkins instance. We recommend the use of a specific service account to trigger the Jenkins job and limit other uses to admin/super user to, for instance, override things if needed. 

\noindent{\bf Cluster IP exposure mitigation.} The main protection from exposing the IP address of an internal \ac{K8s} resource externally is to establish service accounts in the K8s cluster. We recommend assigning specific service accounts to access specific resources within certain namespaces and focus the activities monitoring for malicious behavior to specific users that have access to the internal resources in question.

\noindent{\bf HostPath volume escalation mitigation.} The main protection from the hostPath volume namespace breakout is to restrict the CRUD privileges to higher level accounts. Lower level accounts that needs CRUD privileges must authenticate as a high level user to perform their tasks, which  focuses the activities monitoring to specific limited accounts. We recommend also to acquire dedicated storage so to prevent the need of hostPath volumes being deployed and instead hosting the storage on another machine that would not be part of the main cluster.

\section{Conclusions}\label{sec:conclusions}

We developed a DevOps pipeline case study, and demonstrated four privilege elevation oriented malicious actions against the internal components of the pipeline that show the possibility to maliciously use the system to render a software insecure. The attack scenarios are maliciously abusing the user’s privilege of deploying containers within the K8s cluster, abusing the Jenkins instance to modify files during the CI/CD build phase, modifying the K8s DNS layer to expose an internal IP to external traffic, elevate privileges from a create, read, update, and delete (CRUD) privileges of a low tier account to root privileges. Abiding by the principle of least privilege and ensure lower level accounts do not have any type of admin or root access will assist in reducing the potential attack landscape and allow the security organization to focus on monitoring the activities of limited accounts.



\bibliography{Arxivmain}
\bibliographystyle{ieeetr}

\end{document}